\documentclass[aps,pre,twocolumn,showpacs,floatfix,preprintnumbers,superscriptaddress,amsmath,amssymb]{revtex4-1}
\usepackage{amsmath}
\usepackage{amssymb}
\usepackage{graphicx}
\usepackage{epsfig}
\usepackage{dcolumn}
\usepackage{bm}
\usepackage{array}
\usepackage{mathtools}

\begin{document}
\title{Non-exponential decoherence and subdiffusion in atom-optics kicked rotor}
\author{Sumit Sarkar}
\author{Sanku Paul}
\author{Chetan Vishwakarma}
\author{Sunil Kumar}
\author{Gunjan Verma}
\author{M. Sainath}
\affiliation{Department of Physics, Indian Institute of Science Education and Research, 
Dr. Homi Bhabha Road, Pune 411 008, India}
 \author{Umakant D. Rapol}
 \email{umakant.rapol@iiserpune.ac.in}
\affiliation{Department of Physics, Indian Institute of Science Education and Research, 
Dr. Homi Bhabha Road, Pune 411 008, India}
\affiliation{Center for Energy Science, Indian Institute of Science Education and Research, 
Dr. Homi Bhabha Road, Pune 411 008, India}
\author{M. S. Santhanam}
\affiliation{Department of Physics, Indian Institute of Science Education and Research, 
Dr. Homi Bhabha Road, Pune 411 008, India}

\date{\today}
\begin{abstract}

Quantum systems lose coherence upon interaction with the environment and tend
towards classical states. Quantum coherence is known
to exponentially decay in time so that macroscopic quantum superpositions are
generally unsustainable. In this work, slower than exponential
decay of coherences is experimentally realized in an atom-optics kicked rotor system
subjected to non-stationary L\'{e}vy noise in the applied kick sequence.
The slower coherence decay manifests in the form of quantum subdiffusion that
can be controlled through the L\'{e}vy exponent. The experimental results are
in good agreement with the analytical estimates and numerical simulations for the
mean energy growth and momentum profiles of atom-optics kicked rotor.

\end{abstract}

\maketitle

Quantum systems undergo decoherence due to unavoidable interaction with the environment. As a consequence, 
the macroscopic quantum superpositions are strongly suppressed and classical behaviour emerges 
from the quantum regime \cite{schloss, dec1}. The physics at the borderline of classical
and quantum regimes is not well understood yet and continues to attract
attention \cite{dec1}. Quantum systems coupled to the environment
lose their coherence exponentially fast \cite{zurek}. This is modelled by the decoherence
factor of the form $\exp\left(-t/t_c \right)$, where $t_c$ is the coherence time that 
depends on the system parameters and the strength of coupling to the environment.
In many applications, e.g., in quantum computers and in emerging quantum technologies \cite{bongs}, it 
is necessary to sustain quantum coherences for longer times
which might be possible by tuning $t_c$.

The experimental developments in quantum control and reservoir 
engineering \cite{zoller, kienzler, asjad} in the last two decades have
allowed direct observation of decoherence dynamics. The experiments using
single ions in harmonic traps coupled to an engineered
reservoir of random electric fields \cite{myatt}, dispersively coupled atom and field 
in a cavity \cite{brune} and quantum localized states of ultracold atoms kicked by a noisy 
pulsed optical lattices \cite{ammann,klappauf} have provided evidence for exponential 
decay of quantum coherence. These
experiments allow tuning the coherence time $t_c$ by changing a system
parameter or the coupling to the environment.

An alternative approach \cite{dfs} to prolong $t_c$ is to explore non-exponential or a relatively slow 
coherence decay rates of the form $t^{-\alpha}$, where $\alpha>0$ is the exponent.
Such non-exponential decoherence has been theoretically shown 
for quantum kicked rotor \cite{schomerus1,schomerus2} and
dissipative quantum two-level system \cite{schriefl} influenced by non-stationary noise.
A novel feature in these is that they encompass a regime in which the mean 
coherence time diverges and the quantum system does not complete its transition to the 
classical regime. One challenge in experimental realization of non-exponential decoherence 
is that it will require quantum systems to be sensitive to non-stationary noise within 
the coherence time scales. In this work, experimental realisation of non-exponential
decoherence in the atom-optics kicked rotor (AOKR) system is presented using an unusual form
of non-stationary timing noise.

The standard AOKR system -- cold atoms periodically pulsed (kicked) by the 
electromagnetic fields -- corresponds to a fundamental model of Hamiltonian chaos
\cite{reichl}. In this system, classically chaotic dynamics leads to unbounded, diffusive
mean energy growth whereas its quantized version suppresses the energy growth due to destructive 
quantum interferences. The resulting dynamical localization (DL) is a phase coherent effect
analogous to the Anderson localisation in disordered periodic latices \cite{anderson, andloc, fishman} 
and was experimentally observed in one- \cite{raizen}, two- \cite{delande} and 
three-dimensions \cite{chabe}. With its unambiguous and distinct signatures of energy growth 
in the classical and quantum regimes, AOKR is a suitable test-bed to study decoherence.

In AOKR, DL can be destroyed by inducing decoherence through ({\it i}) spontaneous emission of 
the atoms \cite{ammann, garreau1}, or ({\it ii}) addition of noise in the amplitude of the kicks \cite{klappauf} or in the periodicity of the kick \cite{oskay} or in the phase of the 
periodic kicks \cite{phasenoise}. In contrast to these approaches, we
induce decoherence by suppressing kicks entirely at certain time instants dictated by the 
value of waiting time $\tau$ between successive kicks drawn from
L\'{e}vy distribution $w(\tau)=\alpha \Gamma(\tau) \Gamma(\alpha+1)/\Gamma(\tau+\alpha+1)$,
where $\alpha$ is the L\'{e}vy exponent \cite{schomerus2}. For $0 < \alpha < 1$, this represents a non-stationary
noise with diverging mean waiting time $\langle \tau \rangle$.
We show through experiment and theory that this
scenario leads to non-exponential decoherence rates and manifests as sub-diffusive quantum mean energy growth.
Besides quantum decoherence, these results are relevant in the general context of transport and 
diffusion in chaotic quantum systems \cite{rainer,jwang,sanku} and disordered nonlinear lattices \cite{flach}.

The dimensionless Hamiltonian for AOKR, i.e., two-level atoms in a pulsed standing wave of
near-resonant light, subjected to L\'{e}vy noise is given by
\begin{equation}
H = \frac{p^2}{2} + K \cos x \sum_{n=1}^N ~(1-g_n) ~\phi_{sq}(t-n).
\label{sham}
\end{equation}
where $g_n$ is a telegraph stochastic process that randomly switches
between 0 and 1 \cite{telegraph}. If $g_n=1$, then no kick acts at that instant and if $g_n=0$ then a 
scaled kick strength of amplitude $K$ acts at that time instant. The waiting time between 
the successive occurrences of 0 is taken from $w(\tau)$. The rectangular pulse $\phi_{sq}(.)$ is 
of unit amplitude and $K$
determines whether the dynamics is chaotic, regular or mixed.
The noise-free limit, with $g_n=0$ for all $n$, represents the standard AOKR.
In this limit, the classical system is integrable for $K=0$.
If $0 < K < 1$, the dynamics is mixed as the regular and chaotic regions 
coexist in phase space. When $K > 1$, the system becomes
increasingly chaotic. For $K \approx 5.8$ used in this work, the classical phase space
is largely chaotic over the experimentally accessible energy range.

\begin{figure}[t]
\includegraphics*[width=3.1in]{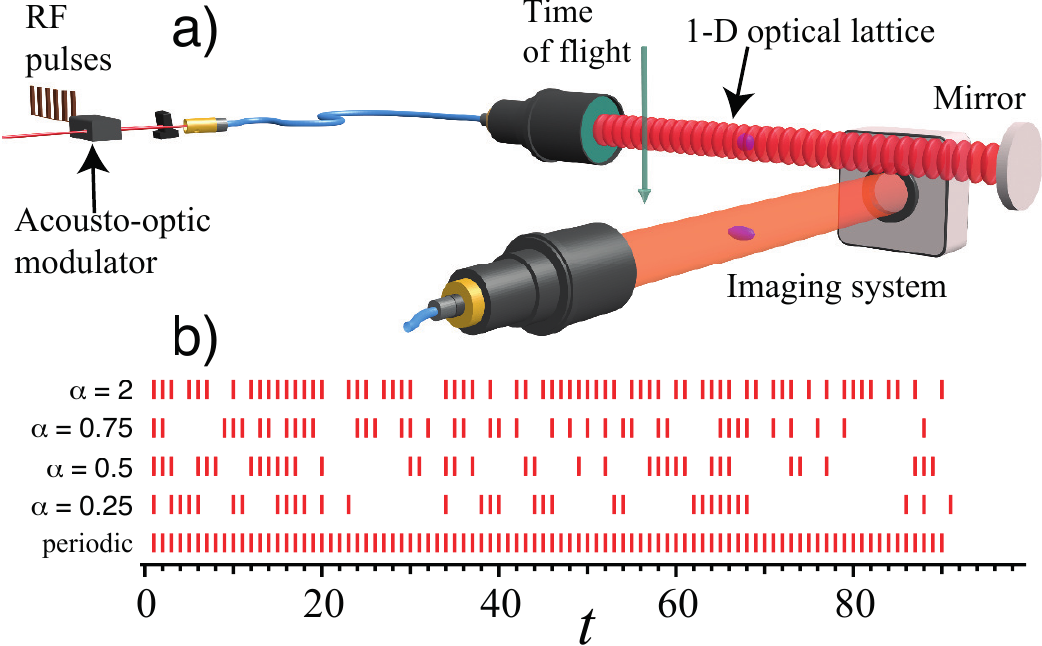}
\caption{(Color online) (a) Experimental schematic shows
the 1-D optical lattice and the absorption imaging system below the lattice.
Off-resonant laser light after passing through an acousto-optic modulator,
single-mode fibre and beam expansion creates a 1-D standing wave.
(b) Position in time axis at which potential kicks are applied
for periodic and various L\'{e}vy exponents $\alpha$.
Kicks are skipped as dictated by the L\'{e}vy distributed waiting times between
successive kicks.}
\label{scheme}
\end{figure}

The experimental set-up to realise the system in Eq. \ref{sham}
is shown in Fig. \ref{scheme}(a).
Approximately $10^7$ atoms of $^{87}$Rb are loaded
into a standard magneto-optic trap and laser cooled to 30 $\mu$K \cite{sunil}. They are then transferred
into a crossed optical dipole trap (wavelength $\lambda=1064$ nm) for further
forced evaporative cooling to 3 $\mu$K.
A far detuned 1-D optical lattice is superimposed on
this cold sample of atoms and pulsed. The lattice laser is $-6.8$ GHz detuned from 
5 $S_{1/2}F=1 \longrightarrow 5 P_{3/2} F=2$ transition of $^{87}$Rb.
For the periodically kicked AOKR, the on-time
of the pulse is 220 ns and the off-time is 10.6 $\mu$s such that the kick period is $T=11.02~\mu$s.
For the experimental parameters used in this work, the kick strength is
$K \approx 5.8$ with $10\%$ uncertainty. The scaled Planck's constant is
$\hbar_s = 8\omega_r T$, where $\omega_r \approx 24$ KHz is the recoil frequency of the lattice beam.
This yields $\hbar_s \approx 2.09$.
Fig. \ref{scheme}(b) displays periodic kick sequence, a crucial ingredient necessary to maintain
the quantum coherence and dynamical localization. We induce decoherence by
using a kick sequence with waiting time, a random integral multiple of $T$, drawn
from L\'{e}vy distribution $w(\tau)$ with exponent $\alpha$. Asymptotically
$w(\tau)$ decays as $\tau^{-\alpha-1}$.
The number of kicks actually
imparted in any finite time interval is a random variable and its mean is
dependent on $\alpha$ (see Fig. \ref{scheme}(b)). This represents the source of noise for the periodically
kicked AOKR.
When $\alpha$ is larger, the number of kicks is larger too.
Properties of L\'{e}vy distribution relevant for our purposes are reviewed in Ref. \cite{schomerus2}.
As shown in Fig. \ref{scheme}(a), the radio-frequency applied to the acousto-optic modulator (AOM)
is pulsed in accordance with $w(\tau)$, thereby pulsing the lattice beam which acts on
the cold atomic cloud during pulse on-times.
The momentum distribution of the cloud is measured after a fixed
time-of-flight by absorption imaging.
The experimental data reported in this paper, for each value of $\alpha$, are averaged over
14 realizations of L\'{e}vy distributed waiting times.

\begin{figure}[t]
\includegraphics*[width=3.3in]{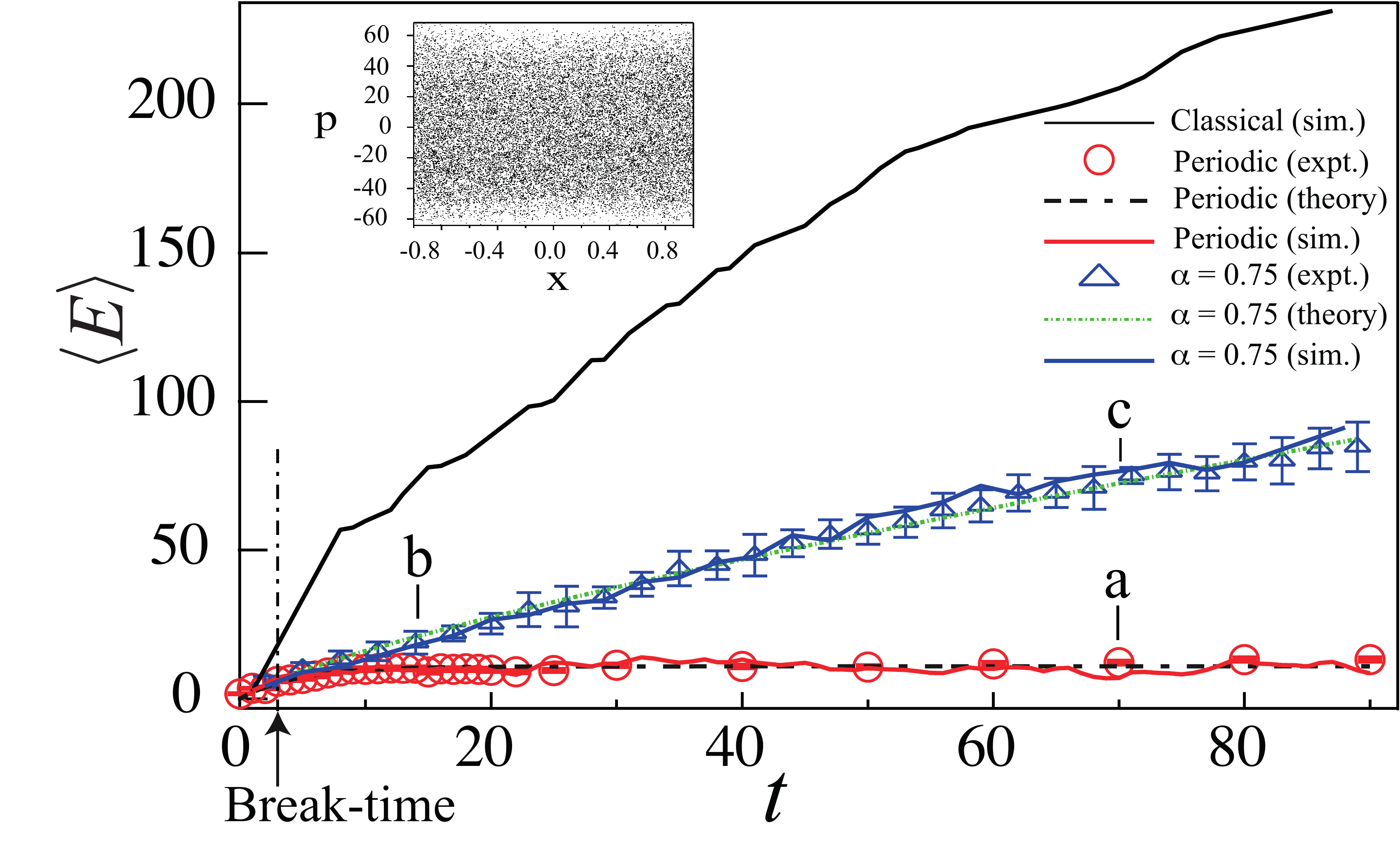}
\caption{(Color online)
Measured mean energy with $K \approx 5.8$ for periodic kick sequence (circles)
and for $\alpha=0.75$ (triangles). Dashed lines are the analytical results and
solid lines are from numerical calculations. Error bars represent the
standard deviation of energy measurement over 14 L\'{e}vy noise realizations.
Black solid line is the simulated
energy growth for classical kicked rotor with $\alpha=0.75$. Inset shows the classical
stroboscopic section for $\alpha=0.75$. Momentum profiles corresponding to labels (a,b,c) are
shown in Fig. \ref{alpha75momprof}. Vertical line marks the break-time $t_b$.}
\label{alpha75}
\end{figure}

\begin{figure*}[t]
\begin{center}
\includegraphics*[width=6.2in]{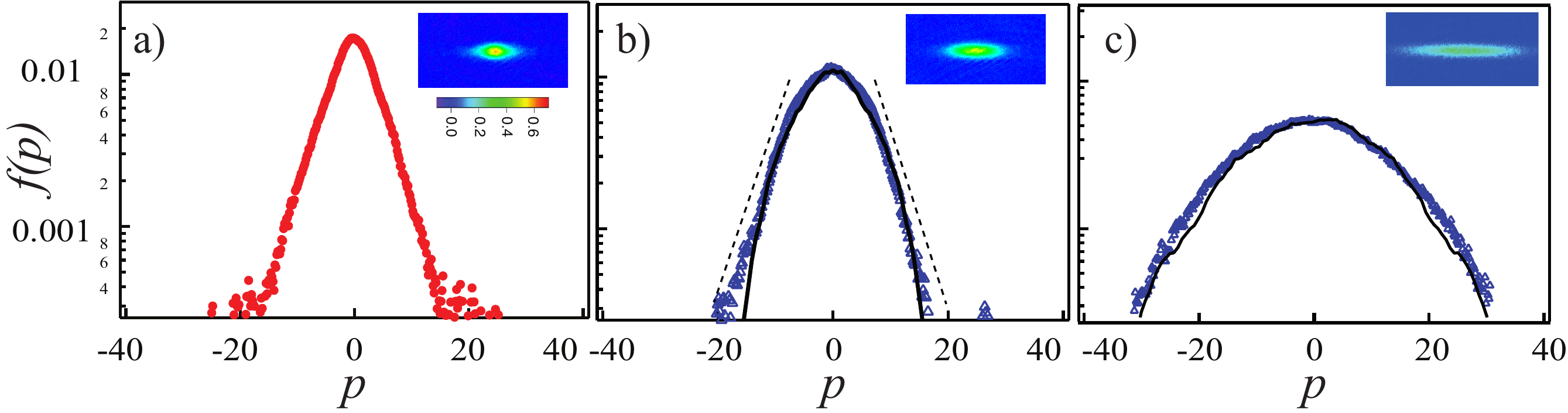}
\end{center}
\caption{(Color online) (a,b,c) Experimentally obtained momentum profiles corresponding
to labels (a,b,c), respectively, in Fig. \ref{alpha75}. The insets show the optical density
of the absorption images from which momentum profiles have been extracted. Solid lines in (b,c)
are obtained from quantum simulation. Dashed lines in (b) are shown as a guide to the eye.}
\label{alpha75momprof}
\end{figure*}

To compare with the experimental results, we performed numerical simulations of 
the noisy AOKR system in Eq. \ref{sham}, by replacing the square pulses with
delta kicks, and the parameters were closely matched with the experimental ones.
The corresponding Floquet operator is repeatedly applied on an initial
state  $\psi(x,0)$, taken to be of Gaussian form.
The waiting time obtained from $w(\tau)$ with exponent $\alpha$ is
incorporated into the Floquet operator.
The simulation results represent an average over 900 realizations of Levy
distributed waiting times.

In Figs. \ref{alpha75} and \ref{alpha75momprof}, we illustrate the central results for 
the case $\alpha = 0.75$.
The classical stroboscopic section (Fig. \ref{alpha75}) obtained by simulations for the
AOKR in Eq. \ref{sham} with L\'{e}vy exponent $\alpha=0.75$ shows that the phase space 
is largely chaotic for momenta up to $p=60$. Corresponding
energy growth displayed as solid (black) line is consistent with approximately quasi-linear
growth proportional to $t$. On the other hand, the measured mean energy $\langle E \rangle$ 
for $\alpha = 0.75$ (see Fig. \ref{alpha75}), averaged over L\'{e}vy noise,
displays sub-diffusive growth. This is in good
agreement with the quantum simulations (blue, solid line) and also with the theoretical
result (green, dashed line) $\langle E \rangle \approx A_0 t + A_1 t^\alpha$. Regression performed on the observed 
$\langle E \rangle$ with this analytical form gives $A_0=0.123$ and $A_1 =2.6(2)$ and $\alpha=0.75(2)$.
For stationary noise imposed on the kicking sequence, exponential decoherence would have resulted in
the experimental data (blue triangles) having similar trend as the classical mean energy growth
(black solid line). At $\alpha=0.75$, the noise is non-stationary and the signature of
slower than exponential decoherence rate comes from this observation that
AOKR with L{\'{e}}vy distributed inter-kick intervals display sub-diffusion rather than normal diffusion
expected under conditions of complete decoherence.
In a related theoretical work on kicked rotor with amplitude noise applied
with L\'{e}vy distributed waiting times \cite{schomerus1, schomerus2, romanelli},
subdiffusion resulting from non-exponential decoherence was noted.
We emphasize that the observed sub-diffusion in this experiment does not arise from
spontaneous emission of photons (at less than 1\% per kick) or
the amplitude noise (about 1\%), both of which are highly suppressed.
This is confirmed by the fact that noise-free AOKR with periodic kicks displays
energy saturation and DL (red circles in Fig. \ref{alpha75}) beyond break-time 
$t_b \approx 3$, in agreement with the established results \cite{raizen, Moore}.

Figure \ref{alpha75momprof}(a,b,c) displays, respectively, the measured momentum profile $f(p)$
corresponding to the labels (a,b,c) indicated in Fig \ref{alpha75}.
For periodically kicked AOKR, $f(p)$ at $t=70 \approx 20 t_b$ obtained from the absorption image shown in 
Figure \ref{alpha75momprof}(a) displays expected exponential profile.
The momentum distribution for $\alpha=0.75$ at $t=14 \approx 4 t_b$ shown in
Fig. \ref{alpha75momprof}(b), extracted from the corresponding absorption image, maintains 
a reasonable exponential profile indicating that coherence is preserved.
However, at $t=70 \approx 20 t_b$ the momentum distribution in Fig. \ref{alpha75momprof}(c) 
is well approximated by 
a Gaussian profile implying loss of coherence. The numerically simulated momentum profile 
is also in good agreement with the measured profiles.

\begin{figure}[t]
\includegraphics*[width=3.4in]{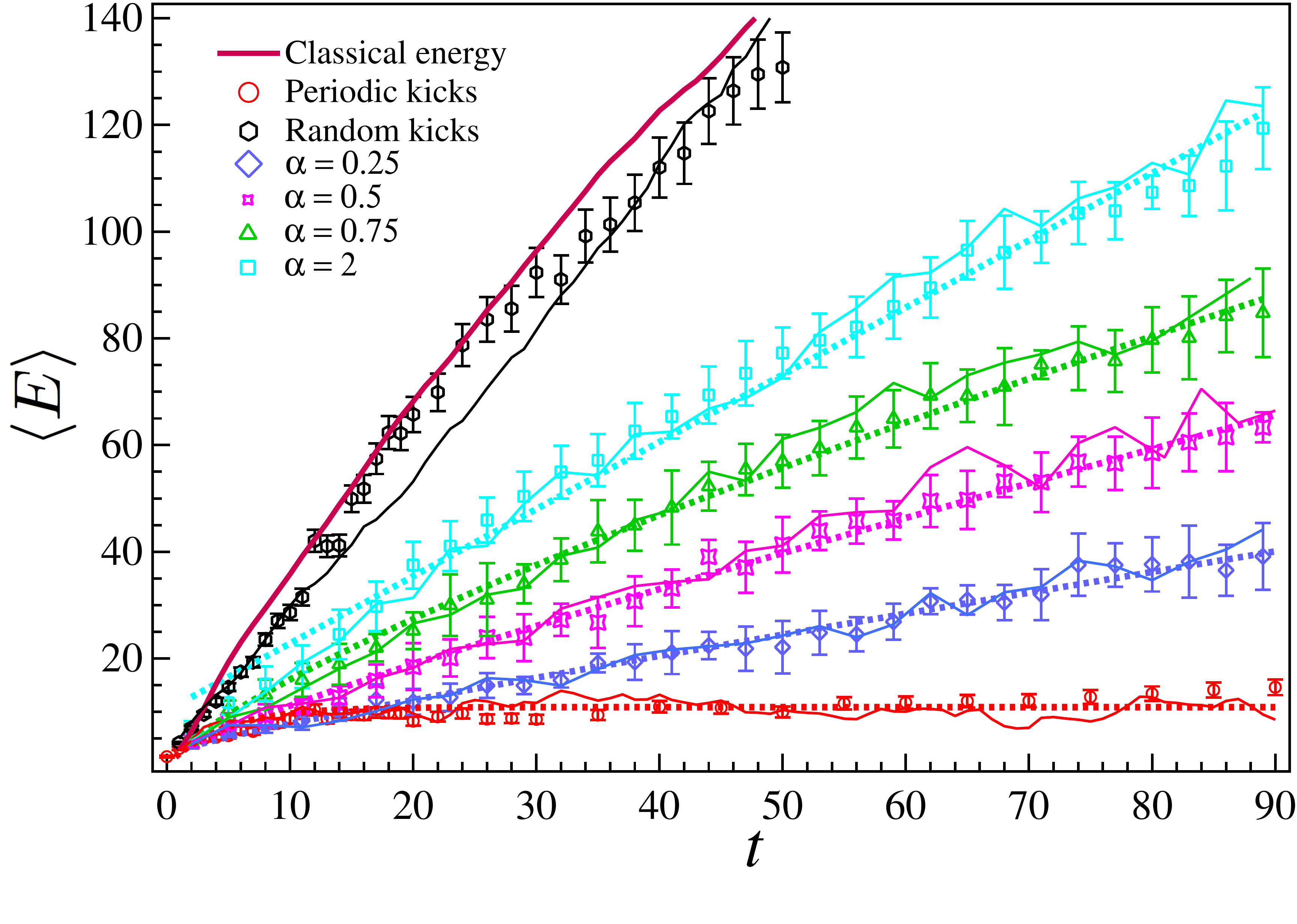}
\vspace{-0.5cm}
\caption{Mean energy growth for various values of L{\'{e}}vi
exponent $\alpha$. Experimental data (symbols) matched against
numerical simulations of AOKR (solid lines) and theoretical
results (dotted lines) in Eqs. \ref{varp1}-\ref{varp2}.}
\label{qdiff}
\end{figure}

The starting point for theoretical analysis is the modified delta-kicked rotor
Hamiltonian of the form
\begin{equation}
H = \frac{p^2}{2} + K \cos x \sum_n (1-g_n) \delta(t-n).
\label{sham1}
\end{equation}
As before, we take the waiting time between successive kicks
from L\'{e}vy distribution $w(\tau)$ characterised by exponent $\alpha$.
The corresponding Floquet operator is
\begin{equation}
\widehat{F}(K) = e^{-\frac{i}{2\hbar_s}p^2} ~ e^{-\frac{i}{\hbar_s} K \cos x} ~ e^{-\frac{i}{\hbar_s} K_n^{'} \cos x},
\label{floq}
\end{equation}
where $K_n^{'}=K g_n$. Following Ref. \cite{cohen} and its extension
to non-perturbative regime in Ref. \cite{schomerus1,schomerus2}, we relate the decoherence factor
to the survival probability of quasi-energy eigenstates $|s\rangle$ of the Floquet operator 
$\widehat{U} = e^{-\frac{i}{2\hbar_s}p^2} ~ e^{-\frac{i}{\hbar_s} K \cos x}$ to
obtain an expression for $\langle p^2 \rangle(t)$. An important ingredient is the use of
random phase approximation to estimate the contribution
$\langle s^{'} | e^{iK_n^{'} \cos x /\hbar_s} | s \rangle$ when the quasi-energy state responds
to noise by transiting from state $|s\rangle$ to $|s{'}\rangle$.
For our noise scenario, we obtain the decoherence factor $D(t,0)$ for the quantum evolution
from $n=0$ until $n=t$ to be,
\begin{equation}
D(t,0) \sim  e^{  -\left( 1 - q^2 \right)t } ~ E_{\alpha} 
     \{ \left( 1 - q^2 \right)  \frac{\sin \pi\alpha}{\pi\alpha} t^{\alpha} \}, ~~~~~~~ (\alpha < 1),
\label{decf1}
\end{equation}
\begin{equation}
D(t,0) \sim  e^{  -(1 - q^2)(1-1/\bar{\tau}) t},
      ~~~~~~~ (\alpha > 1),
\label{decf2}
\end{equation}
where $\bar{\tau}$ is the mean waiting time between kicks, the function 
$q \left(K_t^{'}/\hbar_s \right)$ is given by,
\begin{equation}
q \left(K_t^{'}/\hbar_s \right) = 1 - \frac{K'^2}{2\hbar_s^2} \overline{\cos^2 x} + 
                                    \frac{K'^4}{4! \hbar_s^4} \overline{\cos^4 x} + \dots\dots,
\end{equation}
and $E_{\alpha}(.)$ is the Mittag-Leffler function \cite{abs}.
The decoherence factor is non-exponential for $\alpha< 1$ and is exponential for $\alpha> 1$.
Using the results in Eqs. \ref{decf1}-\ref{decf2}, for $\alpha <1$ and $t>>1$, we obtain the mean energy growth as
\begin{equation}
\label{varp1}
\langle E \rangle \approx A_0 t + A_1 t^\alpha,
\end{equation}
and for $\alpha > 1$ and $t >> 1$, we get
\begin{equation}
\label{varp2}
\langle E \rangle \approx A_2 t.
\end{equation}
where the constants $A_0, A_1$ and $A_2$ depend on the break-time $t_b$ for the corresponding
standard kicked rotor system, $\hbar_s$ and $\alpha$.
These results are in good agreement with the experimental data and simulations presented in Figs. \ref{alpha75}
and \ref{qdiff}.
In order to compare the analytical results for $\langle E \rangle$
with the experimental data, we use $A_0, A_1$ and $A_2$ as fitting parameter.
In Refs. \cite{ammann,klappauf}, sub-diffusive mean energy growth appears as the signature of decoherence
due to spontaneous emission (SE) or amplitude noise (AN). We emphasize that the observed 
sub-diffusion in this work results from non-exponential decohering effect of Levy noise as opposed
to SE or AN (which are in any case suppressed) because
the best fit value of $\alpha$ is close to that used in generating
kick sequences with Levy noise, as predicted by Eqs. \ref{varp1}-\ref{varp2}.
Such physical relevance for the sub-diffusive growth exponent
is absent in case of SE or AN induced decoherence.

A broader picture of the results in Fig. \ref{qdiff} displays the
experimentally measured mean energy growth for values of L\'{e}vy exponent $\alpha=0.25, 
0.5, 0.75$ and 2.0. For comparison, this figure also shows the observed $\langle E \rangle$ 
for AOKR with random kicks or stationary timing noise (STN), i.e, the kick period is $T+\delta$, where 
$-\Delta \le \delta \le \Delta$ is a uniformly distributed random variable with noise 
strength $\Delta \approx 20\%$. The quantum simulations performed for the AOKR system using
experimental parameters agree with the measurements. For $\alpha < 1$, 
subdiffusion is clearly visible, implying non-exponential decoherence. In contrast, for 
STN, approximately normal diffusion is indicative of exponential 
decoherence.  In Fig. \ref{qdiff}, the experimental
results are also compared with the analytical results
in Eq. \ref{varp1}-\ref{varp2} and we obtain a good agreement between the two.
In all these cases, subdiffusion of quantum mean energy growth is highly pronounced for $\alpha < 1$,
the regime in which $\langle \tau \rangle$ diverges as well.
This differs considerably from normal diffusion exhibited by AOKR with STN.
The classical stroboscopic plots corresponding to parameters used in Fig. \ref{qdiff}
have predominantly chaotic features for the energies accessed by these experiments.
It is indeed surprising that the AOKR system 'feels' the L\'{e}vy distributed
waiting times in few tens of kicks and clearly distinguishes it from the case of
STN through different energy growth profiles.

\begin{figure}[t]
\centerline{\includegraphics*[width=3.2in]{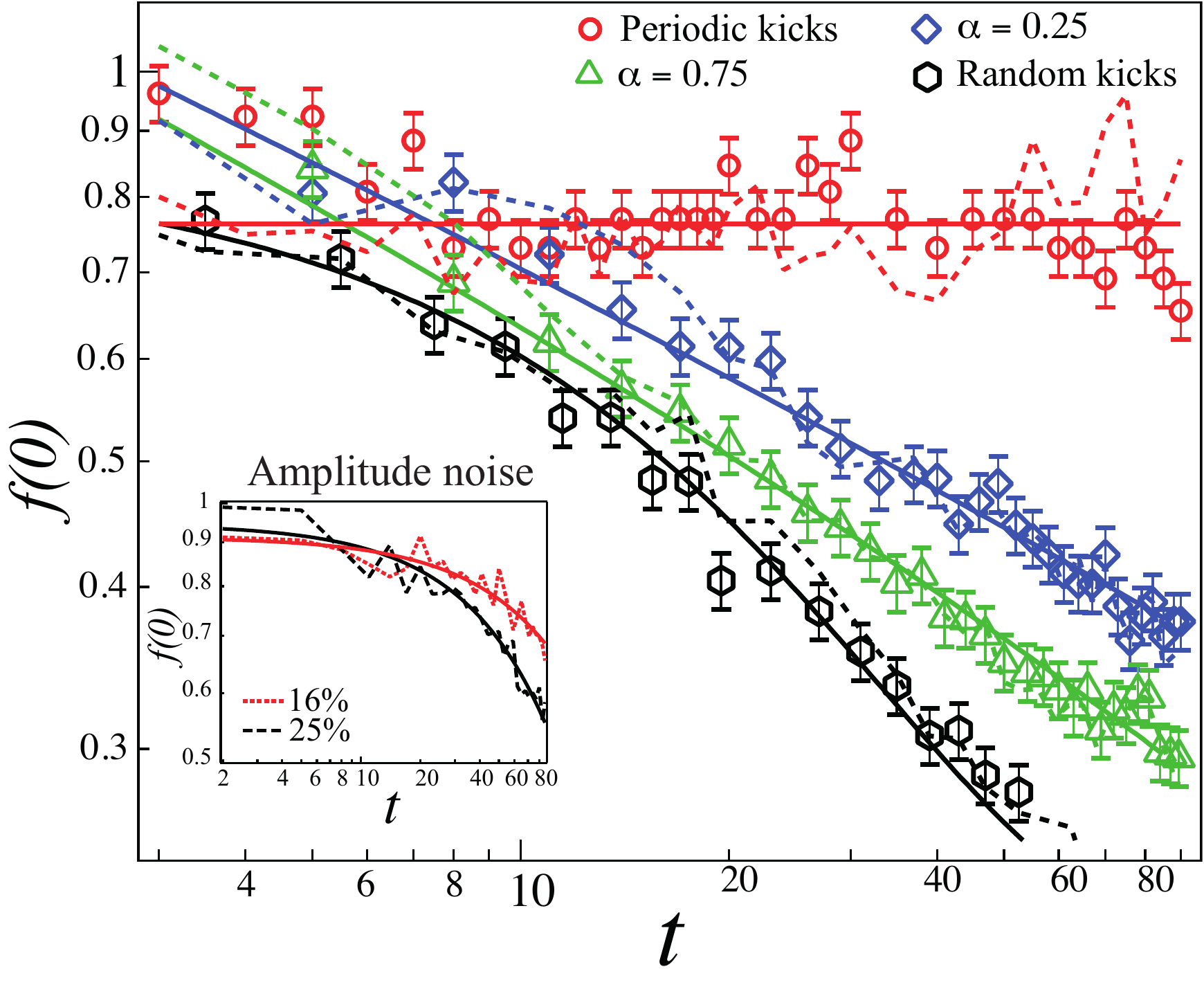}}
\caption{Experimentally measured (symbol) and numerically simulated (dashed line) occupation probability
$f(p=0)$ against time in log-log plot.
(Inset) $f(0)$ for numerically simulated AOKR with stationary amplitude noise
with noise strengths of 16\% and 25\%.
All the solid lines (black, green, blue, red) are to guide the eye.}
\label{momdist}
\end{figure}

The measured momentum profiles provide further evidence 
for slower decoherence  but do not provide
access to phase information.  Instead, decay of occupation probability of zero momentum state
$f(p=0)$ provides an alternative that reflects its coherence \cite{flach1}.
We display measured $f(0)$ in Fig. \ref{momdist} for
various noise scenarios. The decay of $f(0)$ for STN is well represented by an exponential form
while for Levy kicks with $\alpha < 1$ the decay is relatively slow, approximately
linear in log-log plot. In contrast, for noise-free periodically kicked AOKR,
as expected, measured $f(0)$ lacks pronounced decay implying negligible loss of coherence.
The numerical AOKR simulations (dashed lines)
display good agreement with the measurements. Finally, note that
for the decoherence due to {\it stationary} amplitude noise in Ref. \cite{klappauf}, numerically
simulated $f(0)$ displays an exponential decay (inset of Fig. \ref{momdist}) though
associated with subdiffusive growth for $\langle E \rangle$. Thus, inspite of
energy growth being qualitatively similar for stationary and non-stationary noise
in AOKR, they can be distinguished by their qualitatively different coherence decays.

In summary, an experimental realization of slower than exponential decoherence in a noisy
atom-optics kicked rotor system is presented in which the waiting times $\tau$ between
subsequent kicks are chosen from L\'{e}vy distribution characterised
by the exponent $\alpha$. For $0< \alpha < 1$, the noise induced in
AOKR is nonstationary and the accompanying slower
decoherence manifests as subdiffusive quantum mean energy growth.
Remarkably, the AOKR system can 'feel' the non-stationary L\'{e}vy distributed kick
sequence in few tens of kicks. By tuning $\alpha$, mean coherence time 
can be prolonged and it is possible to access the regime of non-exponential 
decoherence rates in experiments. The analytical expressions obtained for subdiffusive 
energy growth are in good agreement with
the experimental and simulation results of AOKR system.

\begin{acknowledgments}
The authors would like to thank the Department of Science and Technology, Govt. of India 
for grants through EMR/2014/000365 and the Nano Mission thematic unit. SP would like to acknowledge 
University Grants Commission, and  SK and MS would like to acknowledge the 
Council of Scientific and Industrial Research, Govt. of India for research fellowship.
The authors thank Dr. Daniel Steck for providing older experimental data related to
amplitude noise that helped strengthen this work.

Author contributions: SS performed the experiment. SP did the analytical and 
numerical simulations. SS, SK and CV
set-up the experiment, SS, MS and GV were involved in data analysis. UDR and MSS conceived
the problem and wrote the paper.
\end{acknowledgments}

\end{document}